\documentstyle[12pt]{article}
\begin{document}
\def\bea{\begin{eqnarray}}
\def\eea{\end{eqnarray}}
\title{\bf { Trace Anomaly and Backreaction of the Dynamical
Casimir Effect }}
\author{
M.R. Setare  \footnote{E-mail: rezakord@yahoo.com}
  \\{Physics Dept. Inst. for Studies in Theo. Physics and
Mathematics(IPM)}\\
{P. O. Box 19395-5531, Tehran, IRAN }\\
 and \\ { Department of Science, Physics group, Kordestan
University, Sanandeg, Iran }\\and \\{Department of Physics, Sharif
University of Technology, Tehran, Iran }}
%\date{\small{\today}}
\maketitle
\begin{abstract}
The Casimir energy for massless scalar field which satisfies
priodic boundary conditions in two-dimensional domain wall
background is calculated by making use of general properties of
renormalized stress-tensor. The line element of domain wall is
time dependent, the trace anomaly which is the nonvanishing
$T^{\mu}_{\mu}$ for a conformally invariant field after
renormalization, represent the back reaction of the dynamical
Casimir effect.

 \end{abstract}
% \begin{document}
\newpage
% \vspace*{10mm}

 \section{Introduction}
In the semiclassical approximation theory of quantum gravity we
are involved
 with calculation the expectation value of energy momentum tensor in special
 vacuum,  \cite{quantum}. However the usual expression
 for the stress tensor includes singular products of the field
 operators for stress tensor.Renormalization theory of the stress
 tensor claims to solve this problem, but it must be mentioned that
 the usual scheme of renormalization includes complexity and somewhat
 ambiguity. For instance, there is no conceptual support for a local
 measure of energy momentum of some given state without any reference
 to any global structure. We know in this frame energy is source of
 gravity and we are not allowed to subtract any unwanted part of
 energy even though it is infinite. So to consider the back reaction effect of the
 quantum field on the gravitational field,  we must
 find a more elaborate renormalization scheme in which the dynamics of
 gravitational field is a vital component.
 In original Casimir effect discovered in 1948 by H.B.G Casimir \cite{casimir}
 we are concerned about force and energy, but we are not usually interested in
 dynamics of the gravitational field. Even in many cases in
 curved boundary  problems, energy is not our main concern.
  Because of unphysical nature of boundary condition
  the energy diverges approaching to curved boundary \cite{deu-can}.
  The Casimir effect can be viewed as a polarization of vacuum by boundary conditions and
  external fields, such as gravitational field. In the present paper we are going
  to consider a simple example in which these two types of sources for vacuum
  polarization are present.There is several
  methods for calculating Casimir energy. For instance,  we can mention mode summation,
  Green's function method \cite{Plunien}, heat kernel method \cite{{kir},{bor}}along with appropriate
   regularization schemes such as point separation \cite{chr},\cite{adler}
  dimensional regularization \cite{deser}, zeta function regularization
  \cite{{haw},{eli1},{eli2}}. But it must be remarked that practically all of the
  methods are successful only for boundary conditions with high
  symmetry in flat space time. In fact we don't have any general procedure for
  renormalizing stress tensor in gravitational background with arbitrary
  boundary ( to see
general new methods to compute renormalized one--loop quantum
energies and energy densities Ref \cite{{gram1},{gram2}}). \\
   In the static situation, the disturbance of the quantum state
  induces vacuum energy  and stress, but no particles are created.
  The creation of particles from the vacuum takes place due to the
  interaction with dynamical external constraints. For example
  the motion of a single reflecting boundary (mirror) can create
  particles \cite{quantum}, the creation of particles by
  time-dependent external gravitational field is another example of
  dynamical external constraints. In two-dimensional space-time and for conformally
   invariant fields the problem with dynamical boundaries can be mapped to
   the corresponding static problem and hence allows a complete study (see
   Refs. \cite{{quantum},{lee}} and references therein).\\
   It has been shown
  \cite{{Nugayev1},{Nugayev2}} that particle creation by black
  hole in four dimension is as a consequence of the Casimir effect
  for spherical shell. It has been shown that the only existence
  of the horizon and of the barrier in the effective potential is
  sufficient to compel the black hole to emit black-body radiation with
  temperature that exactly coincides with the standard result for
  Hawking radiation. In \cite{Nugayev2}, the results
  for the accelerated-mirror have been used to prove above
  statement.\\
  In this paper the Casimir stress tensor for scalar field which satisfies periodic
  boundary conditions in two dimensional
  analog of domain wall space time, is calculated. In this case we
  do not need the boundaries. For the purpose of describing the
  Casimir effect in the one-dimensional cavity by the moving
  mirrors, one can consider a massless field in the
  one-dimensional finite space with two boundaries. The motion of
  the boundaries generally mixes the energy levels of the system.
  However, when the motion of the size of cavity is adiabatic,
  there are no transitions among the energy levels \cite{sass}.
  For this purpose we replace the spatial size of cavity with
  $S^{1}$ space, therefore the scalar field required to satisfy the periodic boundary condition.
  The two dimensional domain wall space time which we have
  considered is the Robertson-Walker type metric, we can regard
  the size of $S^{1}$ as the scale factor of the metric.
  The Casimir stress tensor is obtained by imposing only general
  requirements which is discussed in section 2.We show direct
  relation between trace anomaly and Casimir effect,although we have been aware of
  role of anomalous trace in gravitational background such as
  Hawking effect \cite{chris}. Knowing of Casimir energy in flat space and
   trace anomaly help us to calculate renormalized stress tensor. To see similar
   calculation in background of static domain wall and two-dimensional
   Schwarzschild black hole, refer to \cite{{set},{set1}}, see
   also \cite{vag}for Casimir effect in 2D stringy black hole
   backgrounds.
\section{General properties of stress tensor}
  In semiclassical framework for yielding a sensible theory of back
  reaction Wald \cite{wald} has developed an axiomatic approach.
  There one tries to obtain an expression for the renormalized
  $T_{\mu\nu}$ from the properties (axioms) which it must fulfill.
  The axioms for the renormalized energy momentum tensor are as follow:

  1-For off-diagonal elements standard results should be obtained.

  2-In Minkowski space time standard results should be obtained.

  3-Expectation values of energy momentum are conserved.

  4-Causality holds .

  5-Energy momentum tensor contains no local curvature tensor depending on
  derivatives of the metric higher than second order.

  Two prescriptions that satisfy the first four axioms can differ
  at most by a conserved local curvature term.Wald, \cite{wald2}, showed any prescription
  for renormalized $T_{\mu \nu}$ which is consistent with axioms 1-4 must yield the
  given trace up to the addition of the trace of conserved local curvature .It
  must be noted  (that trace anomalies in stress-tensor,that is,the nonvanishing
  $T^\mu _\mu$ for a conformally invariant field after
  renormalization)
  are originated from some quantum behavior \cite{jac}. In two dimensional
  space time one can show that a trace-free stress tensor can not be
  consistent with conservation and causality
  if particle creation occurs.A trace-free ,conserved stress
  tensor in two dimensions must always remain zero if it is
  initially zero.
  One can show that the "Davies-Fulling-Unruh" \cite{Davies} formula
  for the stress tensor of scalar field which yield an anomalous trace
  ,$T^{\mu} _\mu=\frac{R}{24\pi}$, is the unique one which is
  consistent with the above axioms.
  In four dimensions, just as in two dimensions, a trace-free stress tensor which
  agrees with the formal expression for the matrix elements between orthogonal
  states can not be compatible with both conservation laws and causality .
  It must be noted that, as showed Wald\cite{wald2}, with Hadamard regularization
  in massless case axiom(5) can not be satisfied unless we
  introduce a new fundamental length scale for nature. Regarding all these
   axioms,thus, we are able to get an unambiguous
  prescription for calculation of stress tensor.
  \section{Vacuum expectation values of  stress tensor}

  Vilenkin \cite{vil2} and Isper and Sikvie \cite{ip} have solved
  Einstein's equations in the presence of a planar domain wall by
  approximating the stress energy of the wall as that of an
  infinitely thin plane with positive energy density and negative,
  homogeneouse, and isotropic pressure in the plane of the wall.
  The stress- energy tensor is taken to be
 \begin{equation}
  T^{\nu}_{\mu}=\sigma \delta(z)diag(1,-1,-1,0),
  \end{equation}
  where $\sigma$ is the mass per unite area of the wall and the
  $z$ axis is perpendicular to the wall. In suitable coordinates,
  the metric takes the form
  \begin{equation}
  ds^{2}=e^{-k|z|}[-dt^{2}+dz^{2}+e^{kt} (dx^{2}+dy^{2})].
  \end{equation}
The geometry of hypersurfaces $z=$const, is that of
(2+1)-dimensional de Sitter space. The ($t,z$) part of metric
describes a (1+1)-dimensional Rindler space.\\
Now, just for the sake of simplicity, we consider two dimensions
in which
\begin{equation}
  ds^{2}=-dt^{2}+e^{kt} dx^{2},
  \end{equation}
where we define the above domain wall- type metric on the
space-time $R\times S^{1}$, and $0\leq x \leq l$, where a
dimensional constant $l$ is the standard space size and
$e^{\frac{kt}{2}}$ is the scale factor. We can rewrite the metric
into the conformal flat form by general coordinate transformation
\begin{equation}
  ds^{2}=-dt^{2}+e^{kt} dx^{2}=-c(\eta)(d\eta^{2}-dx^{2}),
  \end{equation}
  where we have introduced a new coordinate $\eta$ such that
  \begin{equation}
  d\eta=e^{-\frac{kt}{2}}dt, \hspace{2cm} c(\eta)=e^{kt}.
  \end{equation}
  From now on, our main goal is to determine a general form of
  conserved energy-momentum tensor, regarding trace anomaly for
  the metric Eq.(3). Once we consider a massless scalar filed in
  the space $S^{1}$, the scalar filed is required to satisfy the periodic boundary condition.
  When we quantize the scalar filed, the conformal anomaly appears
  in general. The quantum effects leads to the motion of scale
  factor \cite{na}, in another word the motion of the scale factor
  is the back reaction of the Casimir effects.\\
   For the non -zero Christoffel symbols of the
  metric Eq.$(4)$we have ;
  \begin {equation}
  \Gamma^{\eta}_{xx}=\Gamma^{x}_{x\eta}=\Gamma ^{x}_{\eta \eta}=\Gamma^{\eta}_{\eta \eta}
  =\frac{\dot{c}}{2c}
   \end{equation}
   Then the conservation equation takes the following form
  \begin{equation}
  \partial_\eta{T^{\eta}_{x}}+\Gamma^{x}_{x \eta}T^{\eta}_{x}-\Gamma^{\eta}_{xx}T^{x}_{\eta}=0
  \end{equation}
  \begin{equation}
  \partial_{\eta}T^{\eta}_{\eta}+\Gamma^{x}_{x\eta}T^{\eta}_{\eta}-\Gamma ^{x}_{x\eta}T^{x}_{x}=0
  \end{equation}
  in which,
  \begin{equation}
  T^{x}_{\eta}=-T^{\eta}_{x}         \hspace{2cm}
  T^{x}_{x}=T^{\beta}_{\beta}-T^{\eta}_{\eta}
  \end{equation}
  and $T^{\beta}_{\beta}$ is anomalous trace in two dimension.Using the
  Eqs. $(6-8)$ it could be shown that
  \begin{equation}
  \frac{d( T^{\eta}_{x}c(\eta))}{d \eta}=0
  \end{equation}
  and
  \begin{equation}
  \frac{d( T^{\eta}_{\eta}c(\eta))}{d
  \eta}=\frac{\dot{c}(\eta)}{2}T^{\beta}_{\beta}.
  \end{equation}
  Then Eq.(10) leads to:
  \begin{equation}
  T^{\eta}_{x}=\alpha'c^{-1}(\eta),
  \end{equation}
  where $\alpha'$ is the constant of integration.The solution of Eq.(11) might be
  written in the following form
  \begin{equation}
  T^{\eta}_{\eta}(\eta)=(H(\eta)+\zeta)c^{-1}(\eta)
  \end{equation}
  where
  \begin{equation}
  H(\eta)=1/2
  \int^{\eta}_{\eta_{0}}T^{\beta}_{\beta}\dot{c}(\eta')d \eta'
  \end{equation}
  and anomalous trace is given by \cite{quantum}
   \begin{equation}
  T^{\beta}_{\beta}=\frac{R}{24\pi}=\frac{k^{2}}{48\pi}.
  \end{equation}
When we chose $\eta_{0}=0$, the function $H(\eta)$ is given by

\begin{equation}
H(\eta)=\frac{k^{2}}{96\pi}(c(\eta)-1).
\end{equation}
Using the Eqs. $(9),(12)$and $(13)$ it can be shown that
   energy momentum tensor takes the following form in $(\eta,x)$coordinates. So
   we have most general form of stress tensor field in our
   interesting background.
   \begin{equation}
   T^\mu\   _\nu(\eta)=\left(\begin{array}{cc}
   H(\eta)c^{-1}(\eta) & 0 \\
   0 & T^{\beta}_{\beta}-c^{-1}(\eta)H(\eta) \
   \end{array}\right)+c^{-1}\left(\begin{array}{cc}
   \zeta& \alpha'  \\
   -\alpha' & -\zeta \
   \end{array}\right)
   \end{equation}

     Now we are going to obtain two constants  $\alpha^{'}$
     and $\zeta$ by imposing the second axiom of
   renormalization scheme. So when we put $k=0$,  we reach the
   special case of flat space-time. The type of boundary
    condition which we choose is periodic boundary condition $\phi(x,t)=\phi(x+2n
    \pi,t)$. This is easiest generalization of Minkowski space
    quantum field theory to non-trivial topological structures in
    a locally flat space-time. This space-time is $R^{1}\times
    S^{1}$, which has two-dimensional Minkowski space line
    element, but the spatial point $x$ and $x+2\pi a$ are
    identified, where $a$ is the radius of circle $S^{1}$. The
    cartesian components of the vacuum expectation values of the
    stress-tensor are as following \cite{quantum}
    \begin{equation}
    \rho =<0|T_{\eta\eta}|0>=\frac{-1}{24\pi a^{2}},
    \end{equation}
    \begin{equation}
    <0|T_{\eta x}|0>=0,
    \end{equation}
   \begin{equation}
   P =<0|T_{xx}|0>=\frac{-1}{24\pi a^{2}}.
   \end{equation}
    Comparing Eqs.(18-20) with Eq. (17) we obtain
    \begin{equation}
    \zeta=\frac{-1}{24\pi a^{2}}, \hspace{2cm} \alpha'=0.
    \end{equation}
Thus we have obtained the energy momentum tensor as direct sum of
two
  terms; first term which present the vacuum polarization in gravitational background in
the absence of boundaries, and second term which come from
periodic boundary condition or in another word due to the
nontrivial topology of space-time. (See the
Ref.\cite{{set},{set1},{vag}} for similar calculation in another
$(1+1)$ dimensional gravitational background).
  \begin{equation}
  <T^\mu  _\nu>=<T^{(g) \mu}  _\nu>+<T^{(t)\mu}  _\nu>
  \end{equation}
  where $<T^{(g) \mu}  _\nu>$ and $<T^{(t)\mu}  _\nu>$ stand for gravitational
 and topological parts respectively. It should be noted that the
 trace anomaly has a contribution just in the first term $<T^{(g) \mu}
 _\nu>$, which comes from the background effect not the
 topological
 one. However, it has a contribution in the total Casimir
 energy-momentum tensor.\\
 The gravitational part pressure are given by
 \begin{equation}
 P^{(g)}=-<T^{(g) 1}  _1>=\frac{-k^{2}}{96\pi}(1+e^{-kt}),
 \end{equation}
 as is clear this pressure come from trace anomaly, represent the back-reactional force
 of the dynamical Casimir effect. The vacuum topological part
 pressure are as following
 \begin{equation}
 P^{(t)}=-<T^{(t) 1}  _1>=\frac{-1}{24\pi a^{2}}e^{-kt}.
 \end{equation}
 Therefore the total pressure naturally contains both the ordinary
 Casimir energy term and the back-reaction term of the dynamical
 Casimir effect.
 \section{Conclusion}
  We have found the renormalized energy-momentum tensor for massless scalar field
  on background of 1+1 dimensional domain wall with periodic conditions, by making
  use of general
  properties of stress tensor only. An essential point of our approach is replacing
  the mirror separation into the size of space $S^{1}$ in the adiabatic
  approximation. The motion of the cavity size is described by varying the radius of  $S^{1}$
  in time. That is, the mirror separation is described by the scale factor of Robertson-Walker
   type metric. The time evolution of the
  scale factor can be regarded as the space-time  $R\times S^{1}$
  with gravity.\\
   The trace anomaly is
  especially important in the special case that the background
  space-time is conformally flat. If the quantum field is also
  conformally invariant, then we have a conformally trivial
  situation. In this case, it turns that anomalous trace
  determines the entire stress-tensor once out the quantum state
  has been specified \cite{{quantum},{br}}.
  We propose that if we know the stress tensor for a given boundary
  in Minkowski space-time, the Casimir effect in gravitational background
  can be calculated. We have found direct relation between trace
  anomaly and total Casimir energy. The conformal anomaly term
  represent the back reaction of the dynamical Casimir effect.
  From the resultant energy-momentum tensor, we have obtained the
  dynamical vacuum pressure. The pressure (dynamical Casimir
  force) includes the back-reactional force of the dynamical
  Casimir effect. The dynamical Casimir force was confirmed to be
  attractive and always stronger than the topological Casimir
  force.

\end{document}